\RequirePackage{amsmath}
\documentclass{article}
\usepackage{spconf,amsmath,graphicx}

\usepackage{amssymb}

\usepackage{enumitem}
\setlist{nosep, leftmargin=14pt}

\usepackage{mwe} 

\usepackage[usenames,dvipsnames]{color}
\usepackage{enumitem}

\title{
Distilling Missing Modality Knowledge from Ultrasound for Endometriosis Diagnosis with Magnetic Resonance Images
}
\name{Yuan Zhang$^{1}$, Hu Wang$^{1}$, David Butler$^{1}$, Minh-Son To$^{2}$, Jodie Avery$^{3}$, M Louise Hull$^{3}$, Gustavo Carneiro$^{4}$}

\address{$^{1}$ Australian Institute for Machine Learning, University of Adelaide, Australia \\
$^{2}$ Flinders Health and Medical Research Institute, Flinders University, Australia \\
$^{3}$ Robinson Research Institute, University of Adelaide, Australia \\
$^{4}$ Centre for Vision, Speech and Signal Processing, University of Surrey, UK}
\begin{document}
%
\maketitle
\begin{abstract}
Endometriosis is a common chronic gynecological disorder that has many characteristics, including
the pouch of Douglas (POD) obliteration, which can be diagnosed using Transvaginal gynecological ultrasound (TVUS) scans and magnetic resonance imaging (MRI).
TVUS and MRI are complementary non-invasive endometriosis diagnosis imaging techniques, but patients are usually not scanned using both modalities and, it is generally more challenging to detect POD obliteration from MRI than TVUS. 
To mitigate this classification imbalance, we propose in this paper a knowledge distillation training algorithm to improve the POD obliteration detection from MRI by leveraging the detection results from unpaired TVUS data.
More specifically, our algorithm pre-trains a teacher model to detect POD obliteration from TVUS data, and it also pre-trains a student model with 3D masked auto-encoder using a large amount of unlabelled pelvic 3D MRI volumes. 
Next, we distill the knowledge from the teacher TVUS POD obliteration detector to train the student MRI model by minimizing a regression loss that approximates the output of the student to the teacher using unpaired TVUS and MRI data.
Experimental results on our endometriosis dataset containing TVUS and MRI data
demonstrate the effectiveness of our method to improve the POD detection accuracy from MRI. 

\end{abstract}
\section{Introduction}
\label{sec:intro}

Endometriosis is a gynecological disorder associated with the growth of endometrial glands and stroma outside the uterine cavity~\cite{lagana2020evaluation, moss2021delayed}. 
The clinical manifestations of endometriosis include infertility and endometriosis-related pain~\cite{vercellini2014endometriosis}. 
As a common chronic gynecological disease, it  affects approximately 1.5 million women worldwide~\cite{doi:10.1056/NEJMra1810764}. There is currently no known way to prevent or cure endometriosis, but early diagnosis, intervention and management may slow or stop the natural disease progression. 
However, the diagnosis of endometriosis can take about 7 years on average after the appearance of initial symptoms~\cite{AIHW_2019}. 
Laparoscopy used to be the diagnostic gold standard~\cite{becker2022eshre}, but with the improvement in the quality and availability of imaging modalities for endometriosis diagnosis, there has been evidence suggesting that accurate endometriosis diagnosis can be achieved with the analysis of TVUS sequences and MRI volumes~\cite{deslandes2020current, indrielle2020diagnostic}.

\begin{figure}[t]
\centering
\includegraphics[width=0.8\linewidth]{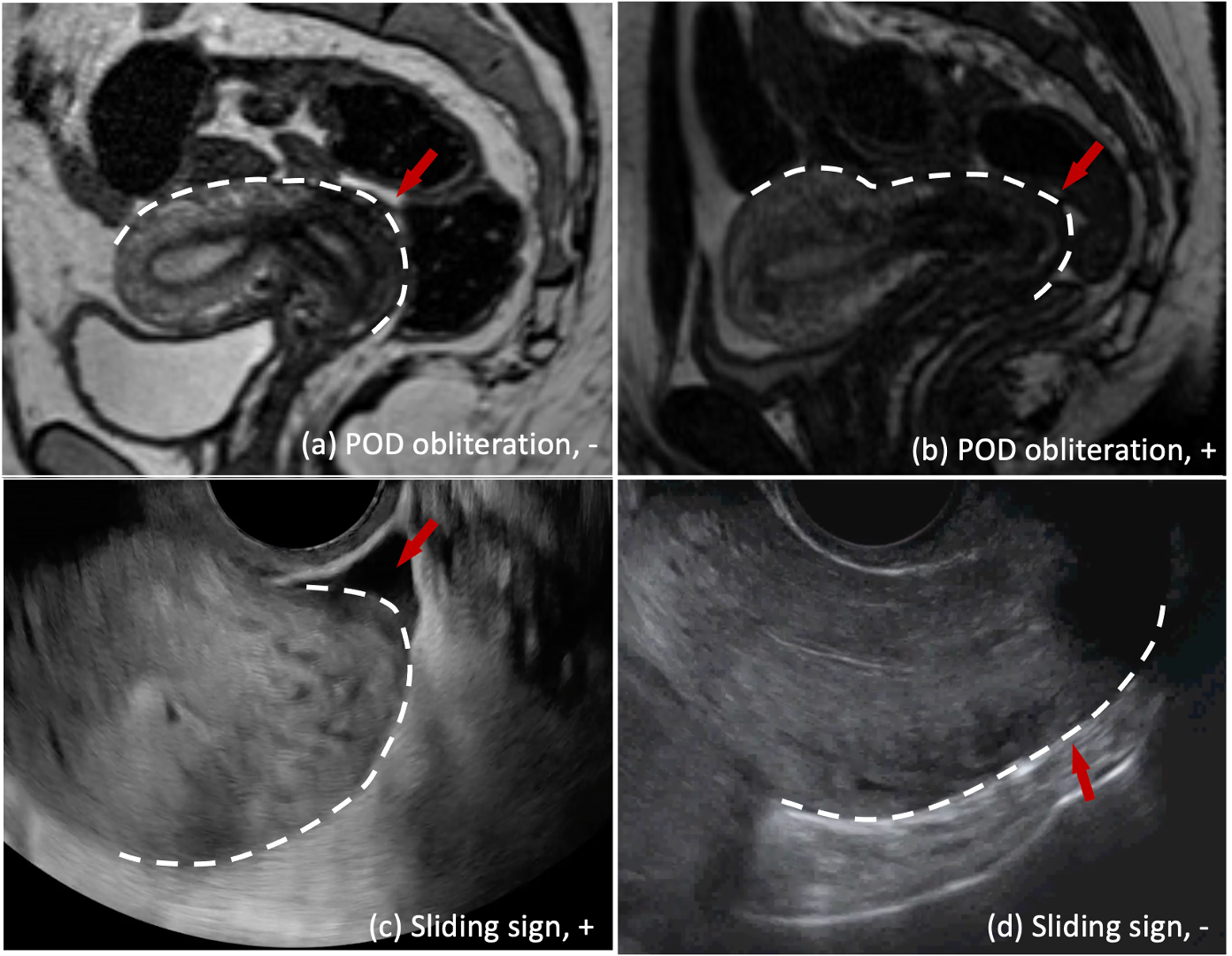}
\caption[]{Examples of POD obliteration on MRI and sliding sign on TVUS. (a) and (b) represent negative and positive POD obliteration sign on sagittal plane MRI, respectively. (c) and (d) represent positive and negative sliding sign on TVUS, respectively. 
} 
\label{fig:POD_example}
\end{figure}

Many of the symptomatic endometriosis cases can be associated with the pouch of Douglas (POD) obliteration, 
which can be diagnosed from complementary TVUS and MRI data, as shown in Fig.~\ref{fig:POD_example}. 
However, in clinical practice, it is difficult to access clinicians who can diagnose endometriosis with one of these modalities, not to mention those who are proficient in both modalities.
For TVUS, POD obliteration can be accurately detected manually~\cite{chiu2019predicting} and automatically~\cite{maicas2021deep} via the ultrasound ‘sliding sign’~\cite{guerriero2016systematic}, which is classified as positive (i.e. normal) or negative (i.e. abnormal), where a negative sliding sign is considered when the anterior rectum or bowel glides cannot freely slide over the posterior upper uterine fundus. 
For MRI, POD can be classified as obliterated or normal, where the
POD obliteration can be characterized by endometrial plaques and dense adhesions between uterosacral ligaments, uterine serosa, ovaries, rectum and vagina on T2-weighted and T1-weighted images~\cite{thalluri2017mri}. 
However, differently from TVUS, the manual POD obliteration detection from MRI can only reach 61.4-71.9\% accuracy~\cite{kataoka2005posterior}. There has been some effort to propose methods that can automatically diagnose deep pelvic endometriosis classification from MRI\footnote{Deep infiltrating endometriosis (DIE) can lead to a partial or complete obliteration of the pouch of Douglas (POD)~\cite{arion2019prediction}.}, 
but we are not aware of methods that can detect POD obliteration from MRI.  

Leveraging the TVUS POD obliteration detection to improve the automated detection accuracy from MRI using an unpaired training set containing scans from both modalities is the main goal of this paper.
We achieve this goal by proposing a new knowledge distillation algorithm based on two stages: 1) pre-training a teacher model to detect POD obliteration from TVUS data, and pre-training a student model with 3D masked auto-encoder (MAE) using a large amount of unlabelled pelvic 3D MRI volumes; and
2) knowledge distillation from the teacher TVUS detector to train the student MRI model by minimizing a regression loss that approximates the output of the student to the teacher using unpaired TVUS and MRI data.
The testing is realised using only MRI data. 
The main innovations of this work are: 
\begin{itemize}
    \item To the best of our knowledge, this is the first POD obliteration detection method that distills knowledge from  TVUS to MRI using unpaired data, with the objective of improving the accuracy of diagnosing endometriosis from MRI;
    \item It is also the first machine learning method that can automatically detect POD obliteration from MRI data with the goal of diagnosing endometriosis.
\end{itemize}
Experimental results on a private endometriosis training set containing unpaired TVUS and MRI data
show the effectiveness of our method to increase the POD detection accuracy from testing MRI volumes. 

\section{Related Work}
\label{sec:format}



\textbf{The automated detection of endometriosis 
from medical imaging} has received some attention lately.
Using ultrasound images, 
Guerriero et al.~\cite{guerriero2021artificial} compared the ability of six machine learning algorithms and neural networks for the diagnosis of endometriosis in the rectosigmoid, where the neural network achieved the highest classification accuracy of 0.73. 
Maicas et al.~\cite{maicas2021deep} constructed a deep learning model based on a temporal residual network to classify POD obliteration from TVUS videos, achieving an AUC of 96.5\% and an accuracy of 88.8\%. 
The methods above use TVUS data alone, but recently, Yang et al.~\cite{yang2021diagnostic} built a bi-model method with one model for MRI and another for TVUS, but they do not explore the relationship between MRI and TVUS, like we propose in this paper.
Furthermore, as mentioned above, it is unlikely that patients will have access to both modalities in clinical practice, which justifies the need for single-modality approaches that have highly accurate endometriosis detection.  

\textbf{Knowledge distillation} 
is a general framework to extract the knowledge learned from a teacher model to a student model by soft-label supervision. 
The original purpose of knowledge distillation is to compress deep learning models, so they can run on resource-constrained devices~\cite{hinton2015distilling}, but in this paper we focus on the transfer of knowledge from a teacher network trained on source modalities to a student network that is trained on different target modalities~\cite{passalis2018learning}.
In medical image analysis, data from different modalities can provide rich and complementary information about diseases, so multimodal knowledge distillation is suitable for scenarios where data or labels for some modalities are missing during training or testing. 
Inspired by knowledge distillation, Dou et al.~\cite{dou2020unpaired} tackled the task of unpaired multi-modal image segmentation. Guan et al.~\cite{guan2021mri} leverage additional supervision distilled from clinical data to improve MRI-based Alzheimer's disease prediction models. Cheng et al.~\cite{chen2021learning} utilise both a pixel-level and an image-level distillation scheme to transfer knowledge from a multimodal-MRI teacher network to a unimodal segmentation student network. However, most unpaired multi-modal studies above focus on MRI and CT scans, which is arguably easier than focusing on MRI and TVUS, which is the case considered in this paper.

\section{Method}
\label{sec:method}

\begin{figure}[t]
\centering
\includegraphics[width=1\linewidth]{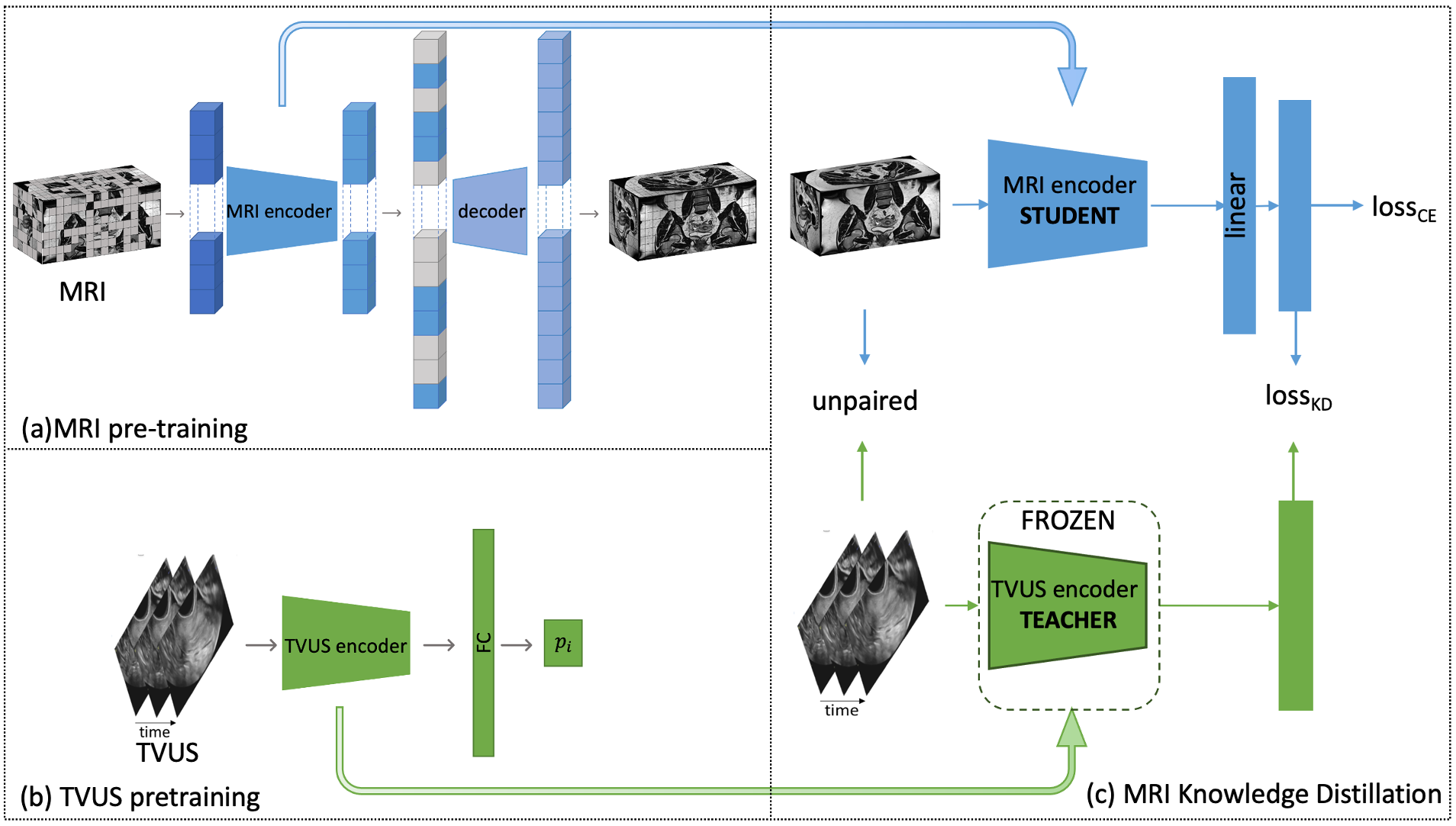}
\caption[]{Proposed POD obliteration detector trained by distilling knowledge to MRI from unpaired TVUS. (a) MRI pre-training with 3D masked auto-encoder, (b) TVUS pre-training with ResNet(2+1)D, (c) MRI Knowledge Distillation from the frozen teacher model pretrained on TVUS.} 
\label{fig:Framework}
\end{figure}

An overview of our proposed TVUS to MRI knowledge distillation model is shown in Figure~\ref{fig:Framework}, which consists of two models: a teacher model pre-trained with a TVUS dataset and a student model pre-trained on an MRI dataset, and then trained on an unpaired dataset of TVUS and MRI data by distilling the knowledge in the representation learned by the teacher model to the student model. 

Formally, let $\mathcal{D}_M=\{ (\mathbf{x}_i, \mathbf{y}_i) \}_{i=1}^{N}$ denote the endometriosis MRI dataset, with $N$ T2 SPC images $\mathbf{x} \in \mathcal{X} \subset  \mathbb{R}^{H \times W \times D}$ and corresponding POD obliteration one-hot label $\mathbf{y} \in  \{0,1\}^2$, where $H$, $W$ and $D$ are height, width and depth of the MRI, respectively. 
For TVUS dataset, let $\mathcal{D}_U^s = \{(\mathbf{x}_i^s, \mathbf{y}_i^s)\}_{i=1}^{M}$ be the video clips dataset, where $\mathbf{x}^s \in \mathcal{X}^s \subset  \mathbb{R}^{H \times W \times T}$ ($H$, $W$ and $T$ are height, width and number of frames), and $\mathbf{y}^s \in \{0, 1\}^2$ denotes the 
POD obliteration one-hot label. 
For the self-supervised pre-training of the MRI POD obliteration detector, we have $\mathcal{D}_{P} = \{ \mathbf{x}_i^p \}_{i=1}^{K}$, which contains $K$ unlabeled  MRI volumes, where the number of unlabeled images is much larger than the labeled images (i.e., $K>>N$ and $K>>M$). 
The teacher model $f_{\theta_{U}}:\mathcal{X}^s \to \Delta$ (with $\Delta \subset [0,1]^2$ being the probability simplex) is trained with dataset $\mathcal{D}^s_U$, the student model $f_{\theta_{M}}:\mathcal{X} \to \Delta$ is pre-trained with dataset $\mathcal{D}_{P}$ and fine-tuned using $\mathcal{D}_M$. The final knowledge distillation model is initialised by the pre-trained student model $f_{\theta_{M}}(.)$, which is then trained from $\mathcal{D}_M$ and $\mathcal{D}^s_U$.
The testing to classify POD obliteration uses the trained $f_{\theta_{M}}(.)$ on the MRI testing images from $\mathcal{D}_M$. 

\subsection{Pre-training}
\label{ssec:pre-train}

For pre-training the MRI encoder, we explore the self-supervised masked auto-encoder (MAE) method~\cite{he2021masked} using the large dataset of un-annotated T2 MRI images $\mathcal{D}_{P}$. 
3D-MAE has an asymmetric encoder-decoder architecture based on 3D Vision Transformer (3D ViT)~\cite{dosovitskiy2020image}, as shown in Figure~\ref{fig:Framework}. During pre-training, each volume is cropped into $8\times8\times8$ patches, then 50\% of the patches are randomly masked. 
The encoder, denoted by $f_{\theta_{E}}:\mathcal{X} \to \mathcal{E}$, takes visible patches embedded by a linear projection 
(of dimension 768) with additional positional embeddings, and processes via the 3D ViT with 12 Transformer blocks whose output is then fed into the decoder, denoted by $f_{\theta_{D}}:\mathcal{D} \to \mathcal{X}$,
together with the masked volume tokens, to reconstruct the original volume at the voxel level. 
After pre-training, the decoder is discarded and the pre-trained encoder is applied to extract MRI features for the downstream classification task. 
During MRI pre-training, we minimize the mean squared error (MSE) on the masked patches between the reconstructed and original volumes in the voxel space, defined as follows:
\begin{equation}
\label{eq1}
\begin{aligned}
    \ell_{MAE}(\mathcal{D}_{P};\theta_{E},\theta_{D})=\frac{1}{L\times K}\sum_{i=1}^{K}  \sum_{l=1}^{L} \| \mathbf{x}_i(l)-\hat{\mathbf{x}}_i(l) \|_2^2,
\end{aligned}
\end{equation}
where $K$ is the size of the unlabelled MRI dataset,
$L$ indicates the number of masked patches, $\mathbf{x}_i(l)$ and $\hat{\mathbf{x}}_i(l)$ represent the voxel values of the $l^{th}$ masked patch in the original and reconstructed volumes, respectively, with the reconstructed volume obtained from $\hat{\mathbf{x}}_i=f_{\theta_{D}}(f_{\theta_{E}}(\mathbf{x}_i))$.

After this self-supervised pre-training, we
take the encoder $f_{\theta_{E}}(.)$, add a linear layer to change the size from 768 (output size of the MRI encoder) to 512 (output size of the TVUS encoder), and add a final classification layer with a 2-dimensional output activated by softmax to form the student network, denoted by $f_{\theta_{M}}(.)$.
We fine-tune all transformer blocks in the encoder and fully connected layer to classify POD obliteration by minimizing the cross-entropy (CE) loss:
\begin{equation}
\label{eq2}
\begin{aligned}
    \ell_{PTM}(\mathcal{D}_M;\theta_{M})=-\sum_{(\mathbf{x}_i,\mathbf{y}_i)\in\mathcal{D}_M} \ell_{CE}(\mathbf{y}_i,f_{\theta_{M}}(\mathbf{x}_i)).
\end{aligned}
\end{equation}

For the TVUS pre-training, we adopted the ResNet (2+1)D model proposed in~\cite{maicas2021deep}. This model contains 18 modules of R(2+1) convolutional layers, with
each convolution being followed by batch normalization.
During this TVUS pre-training, we also minimize the CE loss, as follows:
\begin{equation}
\label{eq3}
\begin{aligned}
    \ell_{PTU}(\mathcal{D}^{s}_{U};\theta_{U})=-\sum_{(\mathbf{x}^{s}_i,\mathbf{y}^{s}_i)\in\mathcal{D}_M} \ell_{CE}(\mathbf{y}^{s}_i,f_{\theta_{U}}(\mathbf{x}^{s}_i)).
\end{aligned}
\end{equation}

\subsection{Knowledge Distillation}
\label{sec:KD}

In the knowledge distillation (KD) stage, 
we consider the pre-trained TVUS model $f_{\theta_{U}}(.)$ as the teacher model, and the pre-trained MRI model $f_{\theta_{M}}(.)$ as the student model.
Given that the pre-trained TVUS model tends to produce superior classification accuracy than the pre-trained MRI model, we aim to fine-tune the MRI model to match the predictions produced by the TVUS model.
The goal is to use this knowledge distillation procedure to improve the classification accuracy of the MRI model, which uses only the MRI volume during testing.
However, recall that we do not have matched TVUS and MRI data for this knowledge distillation procedure, so we match the data based only on their classification labels, i.e., an MRI sample with positive POD obliteration is matched with a random TVUS sample with positive POD obliteration, and similarly for the negative POD obliteration.
Then, the KD training minimises the following loss
\begin{equation}
\label{eq4}
\begin{aligned}
    \ell_{KD}(\mathcal{D}_{M},\mathcal{D}_{U}^{s};\theta_M)=-\sum_{\substack{(\mathbf{x}_i,\mathbf{y}_i)\in\mathcal{D}_M\\(\mathbf{x}^s_j,\mathbf{y}^s_j)\in\mathcal{D}_U^s\\\mathbf{y}_i=\mathbf{y}^s_j}}
    \| f_{\theta_{U}}(\mathbf{x}^s_j) - f_{\theta_{M}}(\mathbf{x}_i)) \|_1.
\end{aligned}
\end{equation}
The loss in~\eqref{eq4} is added to the $\ell_{PTM}(.)$ loss from~\eqref{eq2} to form the final loss that encourages the model to pull the TVUS and the MRI outputs  closer to distill the TVUS classification information to the student network, as follows
\begin{equation}
\label{eq6}
\begin{aligned}
    \ell(\mathcal{D}_{M},\mathcal{D}_{U}^{s};\theta_M)=&\alpha^{epoch} \times \ell_{KD}(\mathcal{D}_{M},\mathcal{D}_{U}^{s};\theta_M)+\\
    &(1-\alpha^{epoch}) \times \ell_{PTM}(\mathcal{D}_M;\theta_{M}),
\end{aligned}
\end{equation}
which is used to estimate the optimal value of $\theta_M$, where $\alpha^{epoch}$ is a parameter to dynamically balance the contributions of the two loss terms during training. 




\section{Experiments}
\label{sec:typestyle}
\subsection{Dataset}
\label{ssec:dataset}

Our private dataset contains: an MRI endometriosis dataset, a TVUS endometriosis  dataset, and a female pelvic MRI  dataset for self-supervised pre-training.

The MRI endometriosis dataset contains 88 T2 SPACE MRI scans from women aged 18-45, including 19 POD obliteration cases. 
These examinations were performed in several clinical sites in Australia.
The scans contain the whole pelvic area, but we focus on the region around the uterus that can display the POD obliteration.
We performed 3D contrast limited adaptive histogram equalization (CLAHE) to improve the local contrast and enhance the definitions of edges in each region of a sequence. 
For the experiments, we use stratified random sampling to split the dataset into 5 folds, each with 17-18 subjects, with 4 folds used for training and 1 for testing, which are rotated to produce a cross-validated result.

The TVUS endometriosis dataset has 749 video clips of the 'sliding sign', including 103 negative (POD obliterated) cases. We follow the data preparation and pre-processing protocol as well as model parameter settings proposed in ~\cite{maicas2021deep} to pre-train the TVUS model. For the knowledge distillation phase, we divide the dataset into negative and positive groups, then use stratified random sampling to split each group into 5 folds and use 4 folds in each group as the training set. 

The female pelvic MRI dataset contains 8,984 volumes. 
In the context of endometriosis research, we accept all scans when the patient is aged between 18 to 45, the physical examination site is pelvis, and the sequence description contains 'T2'. 
It is worth noting that most of these volumes were scanned in different settings, so they may contain signs of other diseases and the scanned area may or may not overlap with the diagnostic area of endometriosis. 
We re-sampled all scans by SimpleITK with an output spacing of $1\times1\times1mm$, and filtered out volumes if the number of slices in any dimension is less than 65. 

\subsection{Training and Evaluation}

In the pre-training phase, the 3D MAE network is trained for 200 epochs on the female pelvic MRI dataset using a batch size of 3 with AdamW optimizer and a learning rate of 1e-3. Then we fine-tuned the pre-trained checkpoint saved from epoch 195 for 25 epochs on the endometriosis MRI dataset with a “5-fold cross validation” strategy. The ResNet(2+1)D network was pre-trained on the Kinetics-400 dataset then fine-tuned for 30 epochs on the TVUS endometriosis dataset using a batch size of 30 with Adam optimizer and a learning rate of 1e-5. For knowledge distillation, we loaded the weights of the $10^{th}$ checkpoint fine-tuned from each fold as the MRI feature extractor of the MRI model and the 
pre-trained TVUS feature extractor. The knowledge distillation network is trained for 10 epochs on the endometriosis MRI dataset using a batch size of 7 with AdamW optimizer and a learning rate of 1e-3 with the same “5-fold cross validation” strategy. We set the hyper-parameter $\alpha=0.85$ based on the cross-validation results.
We evaluate our method by computing Area Under the Receiver Operating Characteristic Curve (ROC AUC) 
for five folds and calculate the mean and standard deviation of them.


\subsection{Results}
\label{sec:results}


The classification results with our proposed method is shown in Table~\ref{table:result}. The ResNet(2+1)D model shows an outstanding classification AUC of 96.9\% from TVUS data. The small amount of training samples limited the generalisation of 3D ViT to classify POD obliteration from MRI volumes with an AUC of 65.0\%, but the MAE pre-training (PT) partially mitigates the issue, improving the AUC to 87.2\%. 
With knowledge distillation (KD), we observe that training a 3D ViT from scratch on such a small dataset is still challenging, with an AUC of 66.7\%. Also, the KD performance of 3D ViT with MAE pre-training (PT) reaches AUC=77.2\%), which is worse than without KD, with AUC=87.2\%, which may be due to the excessive domain shift 
between the pre-training dataset and TVUS dataset. 
By fine-tuning (FT) the model from MAE pre-training, the model improves accuracy from AUC=87.2\% (without KD and FT) to AUC=90.6\% (with KD and FT). 

\begin{table}[h]
\caption{\label{table:result} POD obliteration classification results. 
} 
\scalebox{0.8}{
\begin{tabular}{l|l|l|l}
\hline
\textbf{Method}                                                                                            & \textbf{Training} & \textbf{Testing} &
\textbf{AUC}\\

& \textbf{Modality} & \textbf{Modality} & \textbf{mean$\pm$stddev} \\ \hline \hline 
ResNet(2+1)D                                                                                               & TVUS  &      TVUS       & \textbf{0.969$\pm$0.012}            \\ \hline \hline 
3D ViT                                                                                                     & MRI  & MRI              & 0.650$\pm$0.102            \\ \hline
\begin{tabular}[c]{@{}l@{}}3D ViT: MAE PT\end{tabular}                                     & MRI & MRI              & 0.872$\pm$0.094            \\ \hline
\begin{tabular}[c]{@{}l@{}}3D ViT: KD \end{tabular}                              & MRI,TVUS      & MRI & 0.667$\pm$0.107            \\ \hline
\begin{tabular}[c]{@{}l@{}} 3D ViT: MAE PT + KD \end{tabular}                         & MRI,TVUS      & MRI & 0.772$\pm$0.087            \\ \hline
\begin{tabular}[c]{@{}l@{}}3D ViT: MAE PT + KD + FT  \end{tabular} & MRI,TVUS      & MRI & \textbf{0.906$\pm$0.099}   \\ \hline
\end{tabular}
}
\end{table}



\section{Conclusion}
\label{sec:conclu}

In this paper, we proposed a two-stage algorithm to distill the knowledge from a TVUS to an MRI classifier, thereby improving the POD obliteration classification accuracy of the MRI classifier. 
Through the MAE pre-training, knowledge distillation and fine-tuning, we are able to significantly reduce the distance between the two domains and accomplish a promising knowledge distillation from TVUS to MRI. The efficacy and superiority of our proposed approach are demonstrated by experimental results on our endometriosis datasets. 
In the future, we will introduce a missing modality deep learning approach and expand our proposed method to perform weakly-supervised lesion segmentation, thereby improving the interpretability of the model, so it can be widely applied in future clinical trials.

\section{Compliance with ethical standards}
\label{sec:ethics}

This study was performed in line with the principles of the Declaration of Helsinki. Approval was granted by Human Research Ethics Committee (HREC) of University of Adelaide(Date 01-03-2020/No. H-2020-051) and the Southern Adelaide Clinical Human Research Ethics Committee (SAC HREC) (Date 01-11-2021/No. 111.20).

\section{Acknowledgments}
\label{sec:acknowledgments}

This work received funding from the Australian Government through the Medical Research Futures Fund: Primary Health Care Research Data Infrastructure Grant 2020 and from Endometriosis Australia. 



\bibliographystyle{IEEEbib}
\bibliography{strings,refs}

\end{document}